\documentclass[twocolumn,final,aps,prl,showpacs,superscriptaddress,amsmath,amssymb,amsfonts,floatfix]{revtex4-1}
\usepackage{graphicx}
\usepackage{multirow}
\usepackage{epsfig}

\newcommand{\volb}{Cu$_3$V$_2$O$_7$(OH)$_2\cdot$2H$_2$O}
\newcommand{\newmodel}{$J$-$J'$-$J_1$-$J_2$}
\newcommand{\eff}{\mathrm{eff}}
\newcommand{\sat}{\mathrm{sat}}

\newcommand{\rv}{{\bm{r}}}
\newcommand{\kv}{{\bm{k}}}
\newcommand{\uv}{\bm{u}}
\newcommand{\vv}{\bm{v}}
\newcommand{\Sv}{\bm{S}}
\newcommand{\Tv}{\bm{T}}
\newcommand{\ua}{\uparrow}
\newcommand{\da}{\downarrow}
\newcommand{\Jcal}{{\cal J}}
\newcommand{\Tt}{\tilde{T}}

\newcommand{\bra}[1]{\langle #1|}
\newcommand{\ket}[1]{|#1 \rangle}

\usepackage{url}
\usepackage{color}
\usepackage[colorlinks,breaklinks,bookmarks=true,citecolor=blue,linkcolor=red,urlcolor=blue]{hyperref}
\definecolor{darkred}{rgb}{0.7,0.0,0}

\begin{document}

\title{Magnetic Behavior of Volborthite
Cu$_3$V$_2$O$_7$(OH)$_2\cdot$2H$_2$O\\
Determined by Coupled Trimers Rather than Frustrated Chains} 

\author{O. Janson}
\email{olegjanson@gmail.com}
\affiliation{Institut f\"{u}r Festk\"{o}rperphysik, TU Wien, Wiedner
Hauptstra{\ss}e 8-10, 1040 Vienna, Austria}

\author{S. Furukawa}
\affiliation{Department of Physics, University of Tokyo, 7-3-1
Hongo, Bunkyo-ku, Tokyo 113-0033, Japan}

\author{T. Momoi}
\affiliation{Condensed Matter Theory Laboratory, RIKEN, Wako,
Saitama 351-0198, Japan}
\affiliation{RIKEN Center for Emergent Matter Science (CEMS), Wako,
Saitama 351-0198, Japan}

\author{P. Sindzingre}
\affiliation{Laboratoire de Physique Th\'eorique de la
Mati\`ere Condens\'ee,\ Univ. P. \& M. Curie, 75252 Paris, France}

\author{J. Richter}
\affiliation{Institut f\"{u}r Theoretische Physik, Universit\"{a}t
Magdeburg, D-39016 Magdeburg, Germany}

\author{K. Held}
\affiliation{Institut f\"{u}r Festk\"{o}rperphysik, TU Wien, Wiedner
Hauptstra{\ss}e 8-10, 1040 Vienna, Austria}

\date{\today}

\begin{abstract}
Motivated by recent experiments on volborthite single crystals showing a wide
$\frac13$-magnetization plateau, we perform microscopic modeling by means of
density functional theory (DFT) with the single-crystal structural data as a
starting point. Using DFT+$U$, we find four leading magnetic exchanges:
antiferromagnetic $J$ and $J_2$, as well as ferromagnetic $J'$ and $J_1$.
Simulations of the derived spin Hamiltonian show good agreement with the experimental low-field magnetic susceptibility and high-field magnetization data.
The $\frac13$-plateau phase pertains to polarized magnetic trimers
formed by strong $J$ bonds.  An effective $J\rightarrow\infty$ model shows a
tendency towards condensation of magnon bound states preceding the plateau
phase.
\end{abstract}


\maketitle
The perplexing connection between quantum magnetism and topological states of
matter renewed interest in frustrated spin systems~\cite{kagome:balents10}.  A
prime example is the $S$\,=\,$\frac12$ antiferromagnetic kagome Heisenberg
model (KHM), whose ground state (GS) can be a gapped topological spin liquid,
as suggested by large-scale density-matrix renormalization group (DMRG)
simulations~\cite{kagome:yan11,kagome:depenbrock12}.  Although DMRG results
were recently corroborated by nuclear magnetic resonance (NMR) measurements on
herbertsmithite~\cite{herb:fu15}, alternative methods vouch for a gapless spin
liquid~\cite{kagome:iqbal11b,kagome:iqbal14} and the discussion is still not
settled.

\begin{figure}[tb]
\includegraphics[width=8.6cm]{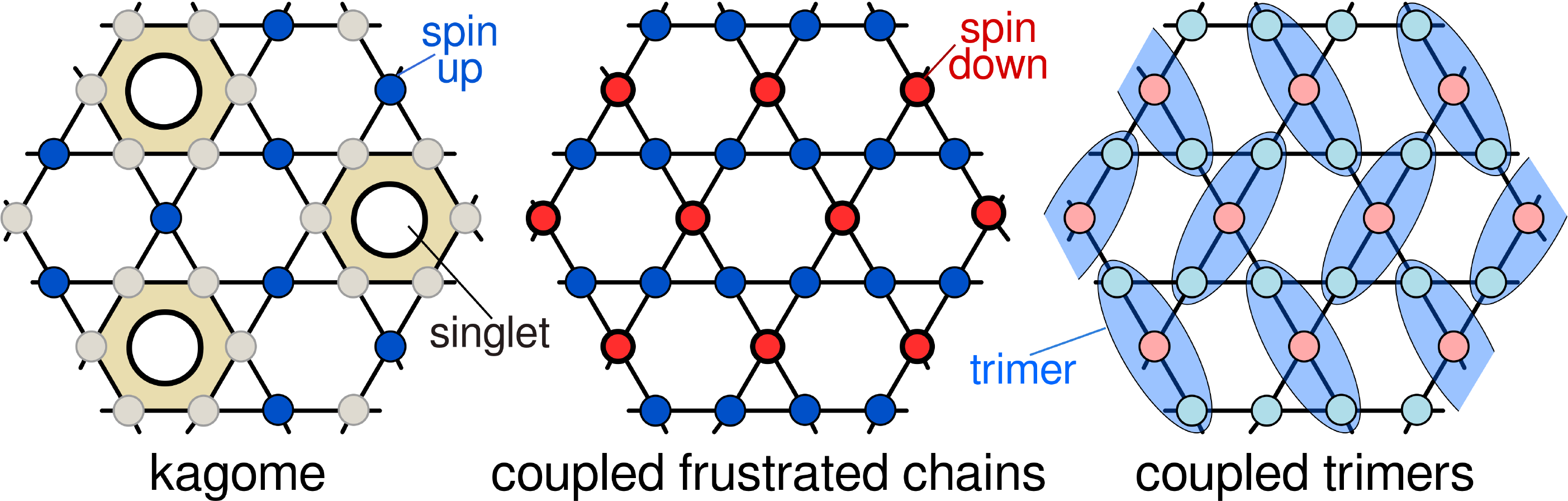}
\caption{\label{fig:plateau}(Color online)
The structure of the $\frac13$-magnetization plateau in the kagome
model (KHM), coupled frustrated chains (CFC) model from
Ref.~\cite{volb:janson10}, and the \newmodel\ model.  }
\end{figure}

One of the remarkable properties of the KHM is the presence of field-induced
gapped phases that manifest themselves as magnetization
plateaus~\cite{kagome:nishimoto13,kagome:capponi13,kagome:schulenburg02}.  A
key ingredient thereof are closed hexagonal loops of the kagome lattice that
underlie the formation of valence-bond solid states~\cite{kagome:capponi13}.
By far widest is the $\frac13$-magnetization plateau, whose structure is
well described by singlets residing on closed hexagons, and polarized spins
(Fig.~\ref{fig:plateau},
left)~\cite{kagome:nishimoto13,kagome:capponi13,kagome:schulenburg02,kagome:cabra05}.  

Despite the considerable progress in understanding both quantum and
topological aspects of the KHM, most theoretical findings still await
their experimental verification.  The reason is the scarceness of
material realizations: only a handful of candidate KHM materials is
known to date.  A prominent example is herbertsmithite, where
$S$\,=\,$\frac12$ spins localized on Cu$^{2+}$ form a regular kagome
lattice~\cite{herb_INS_chiT}.  Other candidate materials feature exchange
couplings beyond KHM as kapellasite~\cite{kapel:janson08, kapel:colman08,
kapel:fak12, kapel:jeschke13, kapel:iqbal15, kapel:bieri15}, haydeeite~\cite{kapel:janson08,
kapel:colman10, hayd:boldrin15, kapel:iqbal15, kapel:bieri15},
francisite~\cite{francs:rousochatzakis15}, or barlowite~\cite{barl:han14,
barl:jeschke15}.

The natural mineral volborthite \volb\ was considered a promising KHM
material~\cite{volb:hiroi01,volb:bert05}, until it was noticed that the local
environment of two crystallographically distinct Cu sites hints at different
magnetically active orbitals~\cite{vesi:okamoto2009}.  Density functional
theory (DFT) calculations show that this has dramatic implications for the spin
physics, giving rise to coupled frustrated chains (CFC) with ferromagnetic (FM)
nearest-neighbor and antiferromagnetic (AF) second-neighbor exchanges, and
interstitial spins that are AF coupled to the two neighboring
chains~\cite{volb:janson10}.  However, detailed structural studies reveal that
below $\sim$300\,K all Cu atoms have the $d_{x^2-y^2}$ as the magnetically active
orbital~\cite{volb:yoshida12ncomm}, questioning the applicability of the CFC
model for volborthite. 
Furthermore, the CFC model features the $\frac13$-magnetization plateau with a
semiclassical ``up-up-down'' structure (Fig.~\ref{fig:plateau}, middle), which
was never observed in powder samples~\cite{volb:yoshida09,
volb:okamoto11}.  Recent magnetization measurements on single crystals
overturned the experimental situation: a broad $\frac13$-magnetization
plateau sets in at $H_{\text{c1}}\!\simeq$\,26\,T and continues up to at least
74\,T~\cite{volb:ishikawa15}.

\begin{figure}[tb]
\includegraphics[width=7.4cm]{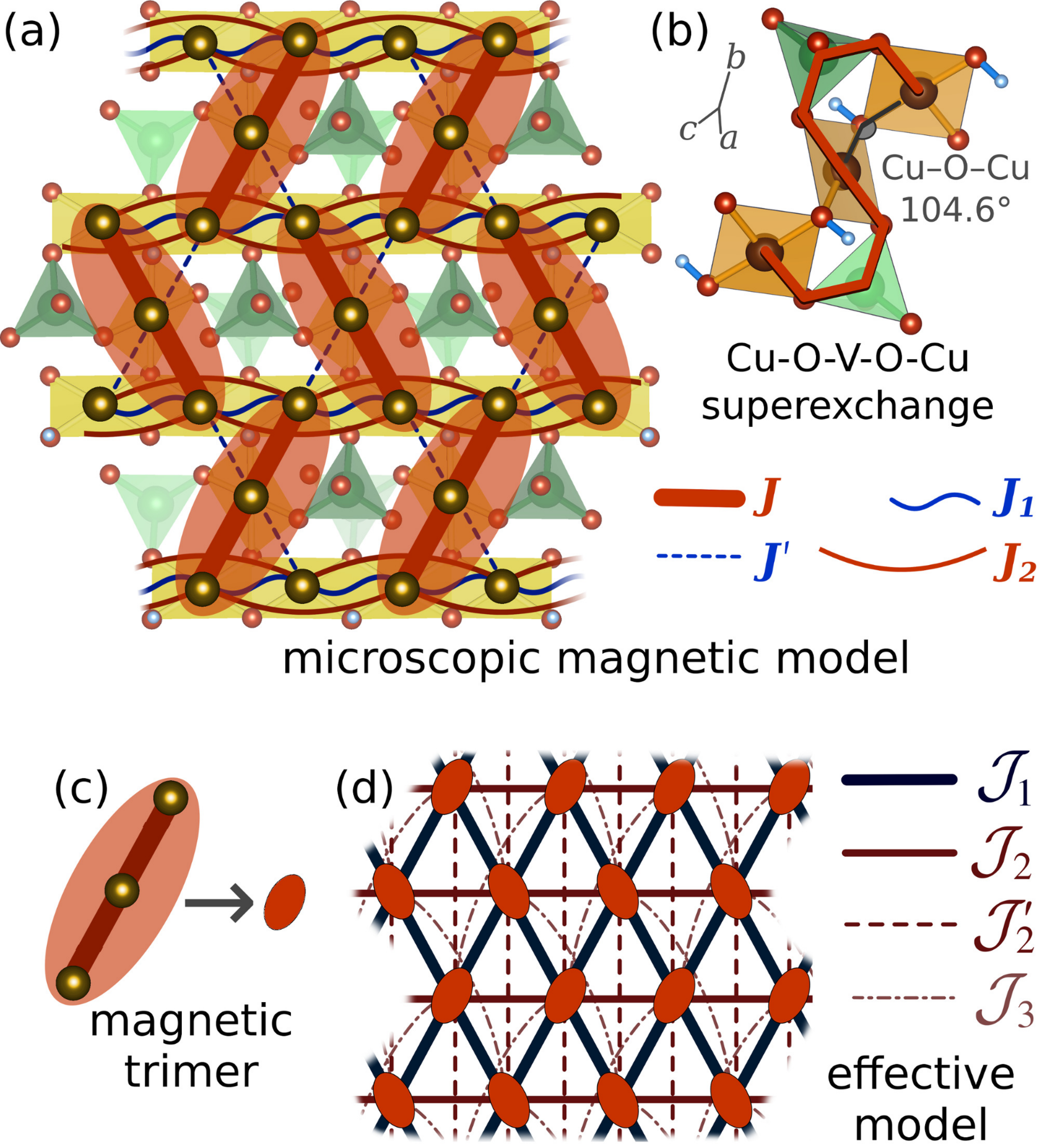}
\caption{\label{fig:model}(Color online) (a) Microscopic magnetic model of
volborthite featuring four relevant exchange couplings: antiferromagnetic $J$
(thick bars) and $J_2$ (solid curved lines), as well as ferromagnetic $J'$
(dashed lines) and $J_1$ (wiggly lines).  Magnetic trimers formed by $J$
exchanges are highlighted (shaded ovals).  Magnetic Cu atoms are shown as large
spheres within CuO$_4$ squares, nonmagnetic V atoms are middle-sized spheres
within VO$_4$ tetrahedra.  (b) The Cu-O-V-O-Cu superexchange paths in the
magnetic trimer. (c) Magnetic trimers form a basis for (d) the effective model
with ferromagnetic $\mathcal{J}_1$, as well as antiferromagnetic
$\mathcal{J}_2$, $\mathcal{J}_2'$, and $\mathcal{J}_3$.
}
\end{figure}

Puzzled by the remarkable difference between the single-crystal and powder
data, we adopt the structural model from Ref.~\cite{volb:ishikawa15} and
perform DFT and DFT+$U$ calculations. We find a microscopic model which is even
more involved than CFC: besides sizable $J_1$ and $J_2$ forming frustrated spin
chains, the coupling between the chain and the interstitial Cu atoms is now
facilitated by two \emph{inequivalent} exchanges, a sizable AF $J$ and a much
weaker FM $J'$.  Due to the dominance of $J$, the magnetic planes break up into
magnetic trimers (Fig.~\ref{fig:model}).  By using exact diagonalization (ED)
of the spin Hamiltonian, 
we demonstrate that this model agrees with the
experimental magnetization data and explains the nature of the plateau phase
(Fig.~\ref{fig:plateau}, right).  Further insight into the low-field and
low-temperature properties of volborthite is provided by analysing effective
models of pseudospin-$\frac12$ moments $T$ living on trimers.  Thus, a model
based on effective exchanges ${\cal J}_1$, ${\cal J}_2$ and ${\cal J}_2'$
supports the presence of a bond nematic phase due to condensation of two-magnon
bound states.  Finally, we conjecture that powder samples of volborthite suffer
from disorder effects pertaining to the stretching distortion of Cu octahedra.

We start our analysis with a careful consideration of the crystal
structure.  Volborthite features a layered structure, with kagome-like planes
that are well separated by water molecules and non-magnetic V$_2$O$_7$ groups.
Magnetic Cu$^{2+}$ atoms within the planes occupy two different sites: Cu(2)
with four short Cu--O bonds forms edge-sharing chains, and interstitial Cu(1)
located in between the chains.  Different structural models in the literature
suggest either squeezed~\cite{volb:basso88,*volb:lafontaine90} or
stretched~\cite{volb:kashaev08,*volb:ishikawa12} Cu(1)O$_6$ octahedra.  The DFT
study of Ref.~\cite{volb:janson10} employed a structure with a squeezed Cu(1)
octahedron.  Although such configuration can be realized at high
temperatures~\cite{volb:yoshida12ncomm}, Cu(1)O$_6$ octahedra are actually
stretched in the temperature range relevant to
magnetism~\cite{volb:yoshida12ncomm, volb:ishikawa15}.  The respective
structural model was never studied with DFT, hence we fill this gap with the
present study.

For DFT calculations~\cite{suppl}, we use the generalized gradient
approximation (GGA)~\cite{PBE96} as implemented in the full-potential code
\textsc{fplo}9.07-41~\cite{FPLO}.  We start with a critical
examination of all structural models proposed so far, by optimizing the H
coordinates and comparing the total energies.  In this way, we find that the
single crystal structure of Ref.~\cite{volb:ishikawa15} has the lowest total
energy~\cite{suppl}.  All further calculations are done for this structural
data set.

To evaluate the magnetic couplings, we project the relevant GGA bands
onto Cu-centered Wannier functions~\cite{suppl}.  The leading transfer
integrals $t$ ($>$50\,meV) of the resulting one-orbital ($d_{x^2-y^2}$)
model are provided in Table~\ref{tab:tJ}.  Their squared values are
proportional to the AF superexchange, which is usually the leading
contribution to the magnetism. However, such one-orbital model fully
neglects FM contributions that are particularly strong for short-range
couplings ($d_{\text{Cu..Cu}}$\,$\lesssim$\,3\,\r{A}).  Hence, to
evaluate the exchange integrals that comprise AF and FM contributions,
we perform DFT+$U$ calculations for magnetic supercells and map the
total energies onto a Heisenberg model.  These results are summarized in
Table~\ref{tab:tJ}.

\begin{table}[tb]
\caption{\label{tab:tJ}
Direct Cu..Cu distances $d_{\text{Cu..Cu}}$ (in \r{A}), transfer
integrals $t$ (in meV) and exchange integrals $J$ (in K).
GGA+$U$ results are provided for three different values of the
on-site Coulomb repulsion $U_d$.  The two numbers in each entry
pertain to the two structurally inequivalent layers; this
minor layer dependence is ignored in the subsequent analysis.}
\begin{ruledtabular}
\begin{tabular}{l r r r r r}
\multirow{2}{*}{} &
\multicolumn{1}{c}{\multirow{2}{*}{$d_{\text{Cu..Cu}}$}} &
\multicolumn{1}{c}{\multirow{2}{*}{$t$}} &
\multicolumn{3}{c}{$J$ (GGA+$U$)} \\ 
& & & $U_d$\,=\,8.5\,eV & 9.5\,eV & 10.5\,eV  \\ \hline
{$J$}   &  3.053/3.058  & $-191$/$-194$ &
    193/205  &  156/167  &  127/136  \\
{$J'$}  &  3.016/\,3.020  &  $-80$/$-84$     &
 $-29$/$-22$ & $-30$/$-25$ & $-32$\,/$-26$ \\ 
{$J_1$} &  2.922/2.923  &  $-98$/$-100$ &
 $-65$/$-65$ & $-76$/$-74$ & $-77$\,/$-76$ \\
{$J_2$} &  5.842/5.842  &     64/64     &
      32/31  &   26/22  &   22/21  \\ 
\end{tabular}
\end{ruledtabular}

\end{table}

Prior to discussing the magnetic model, we should note that the structural
model of Ref.~\cite{volb:ishikawa15} implies the presence of two similar,
albeit symmetrically inequivalent magnetic layers, with slightly different
Cu..Cu distances.  Since the respective transfer ($t$) and exchange ($J$)
integrals for both layers are nearly identical (Table~\ref{tab:tJ}), we can
approximately assume that all layers are equal and halve the number of
independent terms in the model.

The resulting four exchanges, $J$, $J'$, $J_1$, and $J_2$ form the 2D
microscopic magnetic model depicted in Fig.~\ref{fig:model}.  This model is
topologically equivalent to the CFC model: it consists of chains with first-
($J_1$) and second-neighbor ($J_2$) couplings and the interstitial Cu atoms
coupled to two neighboring chains.  However, the exchange between the
interstitial spins and the chains is realized by two different terms: a
dominant AF $J$ and much weaker FM $J'$.  This contrasts with the CFC model,
where both exchanges are equivalent ($J$\,=\,$J'$).

From the structural considerations, the difference between $J$ and $J'$ may
seem bewildering, as Cu..Cu distances (Table~\ref{tab:tJ}) and Cu--O--Cu angles
(104.6$^{\circ}$ versus 102.4$^{\circ}$) are very similar.  Indeed, for the
usual Cu--O--Cu path, the superexchange would be only marginally different for
$J$ and $J'$.  The difference originates from the long-range Cu--O--V--O--Cu
path (Fig.~\ref{fig:model}, b) which provides an additional contribution to
$J$, but not $J'$, since the latter lacks a bridging VO$_4$ tetrahedron. It is
known that long-range superexchange involving empty V $d$ states can facilitate
sizable magnetic exchange of up to 300\,K~\cite{dft:moeller2009}.  Hence, it is
the long-range Cu--O--V--O--Cu superexchange that renders $J$ much stronger
than $J'$.

A distinct hierarchy of the exchanges $J$\,$>$\,$|J_1|$\,$>$\,$J_2,J'$ leads to
a simple and instructive physical picture. The dominant exchange $J$ couples
spins into trimers that tile the magnetic layers.  Each trimer is connected to
its four nearest neighbors by FM $J'$ and $J_1$, and to its two
second-neighbors by AF $J_2$ (Fig.~\ref{fig:model}).  In contrast to the CFC
model, where frustration is driven exclusively by $J_2$, the coupled trimer
model has an additional source of frustration: triangular loops formed by $J$,
$J'$ and $J_1$. Together with $J_2$, they act against long-range magnetic
ordering.

\begin{figure}[tb]
\includegraphics[width=8.6cm]{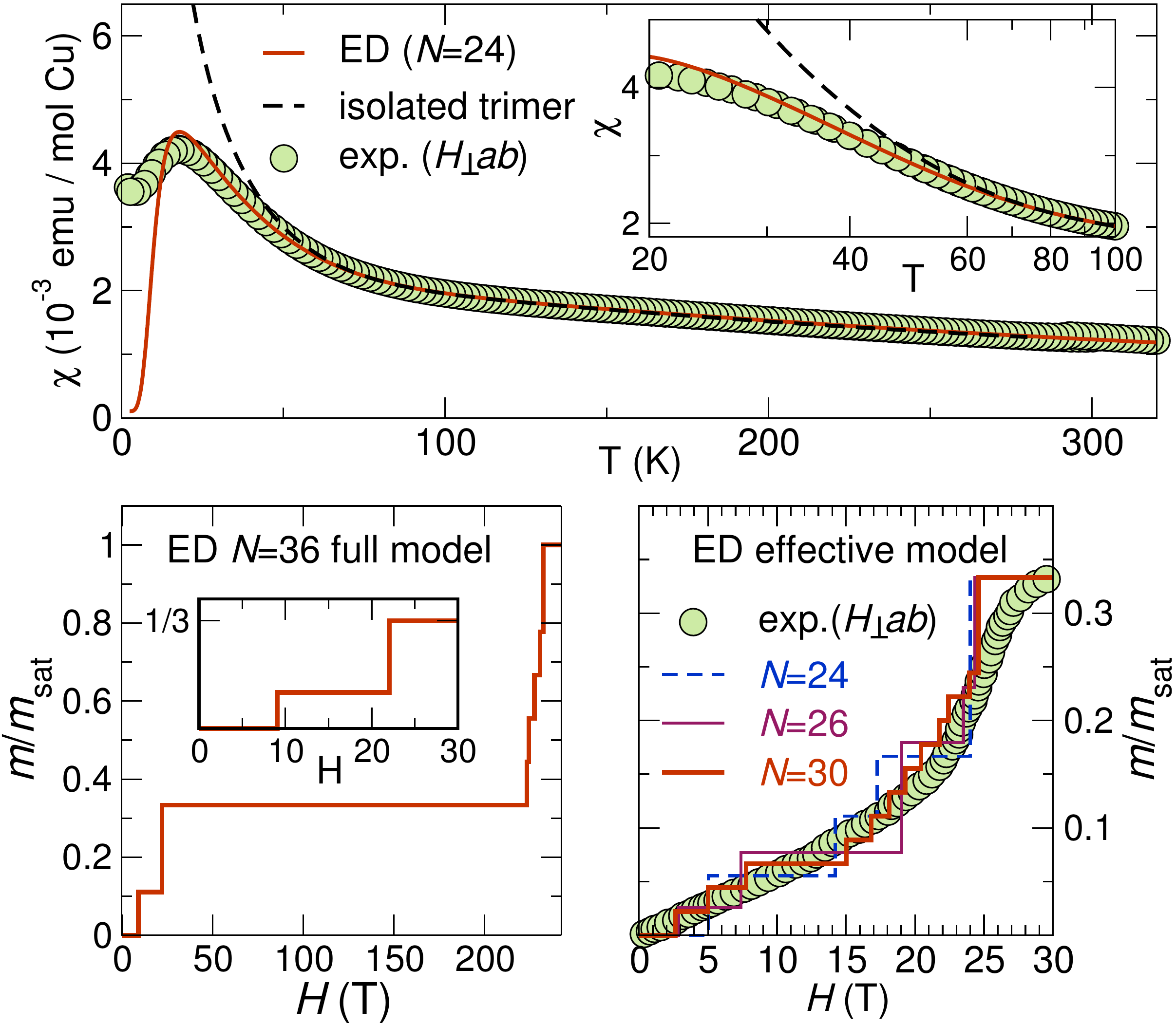}
\caption{\label{fig:simul}(Color online) Top: magnetic
susceptibility of the microscopic spin Hamiltonian calculated by ED on a
$N$\,=\,24 site lattice compared to experiment~(Ref.~\cite{ishikawaXX}) and an
isolated trimer model.  Bottom left: GS magnetization curve simulated on a
lattice of $N$=36 spins for the same model. Insets are magnifications of the
respective data.  Bottom right: GS magnetization of the full effective
model~\cite{suppl} with $N$\,=\,24, 26, and 30 pseudospins compared to
experiment~(Ref.~\cite{volb:ishikawa15}).}
\end{figure}

DFT+$U$-based numerical estimates for the leading exchange couplings allow us
to address the experimental data.  To simulate the temperature dependence of
the magnetic susceptibility $\chi$, ED of the spin Hamiltonian is performed on
lattices of $N$\,=\,24 spins, using the approximate ratios of the exchange
integrals $J$:$J'$:$J_1$:$J_2$\,=\,1:$-0.2$:$-0.5$:0.2 (Table~\ref{tab:tJ}).
The simulated curves are fitted to the experiment by treating the overall
energy scale $J$, the Land{\'e} factor $g$ and the temperature-independent
contribution $\chi_0$ as free parameters.  In this way, we obtain a good fit
down to 35\,K with $J$\,=\,252\,K, $g$\,=\,2.151 and
$\chi_0$\,=\,1.06$\times$10$^{-4}$ \,emu\,/\,$[$mol Cu$]$
(Fig.~\ref{fig:simul}).  ED even reproduces the broad maximum at 18\,K, which
stems from short-range antiferromagnetic correlations.  Deviations at lower
temperatures are finite-size effects.

After establishing good agreement with the $\chi(T)$ data, we employ a larger
lattice of $N$\,=\,36 spins and calculate the GS magnetization curve, which
shows a wide $\frac13$-magnetization plateau between the critical fields
$H_{\text{c1}}$ and  $H_{\text{c2}}$ (Fig.~\ref{fig:simul}, bottom left).
Scaling with $J$ and $g$ from the $\chi(T)$ fit, without any adjustable
parameters, yields $H_{\text{c1}}$\,=\,22\,T in agreement with the experimental
$H_{\text{c1}}$\,=\,26\,T.  In the plateau phase, first- and second-neighbor
spin correlations within each trimer amount to $\langle{\bf S}_0{}\cdot{\bf
S}_1\rangle$\,$\equiv$\,$\langle{\bf S}_1{}\cdot{\bf S}_2\rangle$\,=\,$-0.4938$
and $\langle{\bf S}_0{}\cdot{\bf S}_2\rangle$\,=\,0.2470, very close to the
isolated trimer result ($-\frac12$ and $\frac14$, respectively~\cite{suppl}).
Hence, the $\frac13$-plateau phase can be approximated by a product of
polarized spin trimers formed by strong $J$ bonds (Fig.~\ref{fig:plateau},
right), and thus is very different from the plateau phases of the KHM
(Fig.~\ref{fig:plateau}, left) and the CFC model (Fig.~\ref{fig:plateau},
middle).  The plateau stretches up to a remarkably high
$H_{\text{c2}}$\,$\simeq$\,225\,T, at which the spin trimers break up, allowing
the magnetization to triple.

ED-simulated spin correlations indicate that the simplest effective
model\,---\,the isolated trimer model\,---\,already captures the nature of this
plateau phase.  On general grounds, we can expect the isolated trimer model to
be valid only at high temperatures.  However, it provides a surprisingly good
fit for magnetic susceptibility down to 60\,K (Fig.~\ref{fig:simul}, top),
i.e.\ at a much weaker energy scale than the leading exchange
$J$\,$\simeq$\,250\,K (Fig.~\ref{fig:simul}).  This motivates us to treat the
inter-trimer couplings perturbatively and derive a more elaborate effective
model valid at low temperatures and in low fields ($T,g\mu_{\text{B}}H/k_{\text{B}}\ll J$).

To this end, we adopt the lowest-energy doublet of each trimer at $H$\,=\,0 as
the basis for a pseudospin-$\frac12$ operator $\mathbf{T}_i$.  The
$\frac13$-plateau phase corresponds to the full polarization of
pseudospin-$\frac12$ moments ($m_\sat^\eff$\,=\,$m_\sat/3$).
Degenerate perturbation theory to second order in the inter-trimer couplings
yields an effective Heisenberg model on a triangular lattice with spatially
anisotropic nearest-neighbor couplings ${\cal J}_1$\,=\,$-34.9$\,K and ${\cal
J}_2$\,=\,36.5\,K and much weaker longer-range couplings such as ${\cal
J}_2'$\,=\,6.8\,K and ${\cal J}_3$\,=\,4.6\,K shown in Fig.
\ref{fig:model}~(d)~\cite{suppl}.  The competition between FM ${\cal J}_1$ and
AF ${\cal J}_2$ underlies the frustrated nature of the effective model.  
Larger finite lattices available to ED of the effective model allow us to
amend the critical field $H_\mathrm{c1}$ estimate compared to the full
microscopic model (Fig.~\ref{fig:simul}, bottom left) and reproduce a
pronounced change in the $M(H)$ slope (Fig.~\ref{fig:simul}, bottom right),
which agrees with the experimental kink at $\sim$22\,T~\cite{volb:ishikawa15}.

Effective models provide important insights into the nature of field-induced
states.  Recent NMR experiments on single crystals revealed the emergence of
the incommensurate collinear spin-density-wave (SDW) phase
``II'' ($H$\,$<$\,23\,T) and the ``N'' phase preceding the plateau
(23\,T\,$<$\,$H$\,$<$\,26\,T)~\cite{volb:ishikawa15}.  We first address the
nature of the latter phase, by treating the fully-polarized pseudospin state
(the 1/3-plateau state of volborthite) as the vacuum and analyzing the magnon
instabilities to it.

\begin{figure}[tb]
\includegraphics[width=8.6cm]{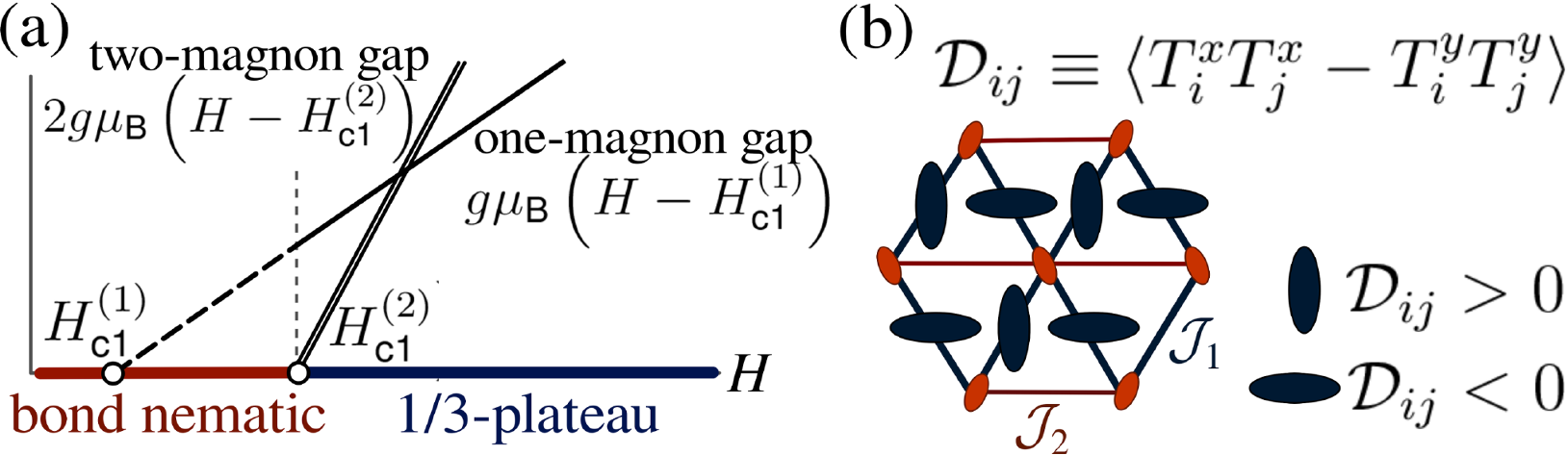}
\caption{\label{fig:nematic}(Color online)
(a) The behavior of one- and two-magnon gaps in the
$\frac{1}{3}$-plateau phase, which gives rise to a bond nematic
phase.  (b) Schematic picture of the bond nematic phase in the
effective model. Orientation of dark ellipses represents the sign of
directors $\mathcal{D}_{ij}$ on ${\cal J}_1$ bonds~\cite{footnote1}. 
}
\end{figure}

To this end, we resort to a model with three leading effective couplings ${\cal
J}_1$, ${\cal J}_2$, and ${\cal J}_2'$.  This model is equivalent to the
frustrated FM square lattice model, where a bond nematic order emerges owing to
condensation of two-magnon bound states (bimagnons) for ${\cal J}_2={\cal
J}_2'\gtrsim 0.4|{\cal J}_1|$~\cite{eff:shannon06}.  Here we take the approximate
ratio ${\cal J}_2/|{\cal J}_1|$\,=\,1 of the perturbative estimates, and study the
influence of ${\cal J}_2'$ on the ground state.  We find that the bond
nematic order is robust for ${\cal J}_2' / |{\cal J}_1|\gtrsim
0.3$~\cite{suppl}, as signaled by the occurrence of bimagnon condensation at
$H_\mathrm{c1}^{(2)}$, at which the plateau state is already destabilized, but
before single-magnon condensation sets in at $H_\mathrm{c1}^{(1)}$
(Fig.~\ref{fig:nematic}, a).  The bond nematic phase shows no long-range
magnetic order besides the field-induced moment, but it is characterized by a
bond order with an alternating sign of directors
$D_{ij}\equiv\langle{}T^x_iT^x_j-T^y_iT^y_j\rangle$ residing on ${\cal J}_1$
bonds (Fig.~\ref{fig:nematic}, b)~\cite{footnote1}.  This phase is a
viable candidate for the experimentally observed ``N'' phase, whose NMR spectra
are not explained by simple magnetic orders~\cite{volb:ishikawa15}.

While bimagnons are stable in a wide region of the ${\cal J}_1$-${\cal
J}_2$-${\cal J}_2'$ model, longer-range effective couplings such as ${\cal
J}_3$ tend to destabilize bimagnons. However, a slight tuning of the
microscopic model (e.g., increasing of $|J'|$) can counteract this effect,
thereby recovering the nematic phase~\cite{suppl}. Long-range effective
couplings are also sensitive to weak long-range exchanges neglected in the full
microscopic model.  In the absence of experimental estimates for these small
exchanges, the ${\cal J}_1$-${\cal J}_2$-${\cal J}_2'$ effective model is an
adequate approximation, which allows us to study the nature of the
field-induced phases in volborthite.

Below 23\,T, NMR spectra indicate the onset of an incommensurate collinear
phase ``II''~\cite{volb:ishikawa15}.  Unfortunately, incommensurate spin
correlations produce irregular finite-size effects that impede an ED
simulation.  Yet, on a qualitative level, further truncation of the model to
the effective couplings ${\cal J}_1$ and ${\cal J}_2$ leads to an
anisotropic triangular model, for which a field-theory analysis predicts the
SDW order for $m\lesssim \frac23 m_\sat^\eff = \frac29
m_\sat$~\cite{eff:starykh10}.

Next, we go a step beyond the Heisenberg model and consider antisymmetric
Dzyaloshinskii-Moriya (DM) components for the leading couplings
$J$ and $J_1$. By performing noncollinear DFT+$U$ calculations with
{\tt vasp}~\cite{VASP, *VASP_2}, we obtain 
$|D_1|/J_1$\,$\simeq$\,0.12 with $\vec{D}_1$ nearly orthogonal to the
frustrated chains.  DM vectors $\vec{D}$ within the trimers are nearly
orthogonal to the respective interatomic vectors and amount to
$|D|$/$J$\,$\simeq$\,0.09~\cite{suppl}.  We analyzed the influence of $D$ for
isolated dimers with ED and found a minute change in spin correlations in the
plateau state, which amounts to 2\,$\%$ at most.  However, these DM
interactions are the leading anisotropy at low fields,
and can give rise to the two consecutive transitions to the incommensurate
phase ``I'' ($T$\,$<$\,1\,K, $H$\,$<$\,4\,T)~\cite{volb:yoshida12ncomm}.

Finally, we address the intriguing question why the $\frac13$ plateau has not
been observed in powder samples.  We remind the reader that the trimers are
underlain by the stretching distortion of Cu(1)O$_6$ octahedra, which selects
two out of four neighboring VO$_4$ octahedra for the $J$ superexchange pathway
(Fig.~\ref{fig:model}, b).  In single crystals, the distortion axes are fixed,
and the trimers form an ordered parquet-like pattern.  Powder samples on the
other hand are more prone to a random choice of the distortion axis.  A single
defect of this type permutes $J$ and $J'$, ruining the trimer picture locally.
This tentative scenario explains the absence of a plateau and the strong
dependence on the sample quality in the powder magnetization data.

In summary, the stretching distortion of the magnetic Cu(1)O$_6$ octahedra in
volborthite leads to the model of coupled trimers, very different from the
anisotropic kagome and coupled frustrated chain models discussed in earlier
studies.  Based on DFT calculations and ED simulations, we conclude that i) the
microscopic magnetic model of volborthite contains four exchanges with a ratio
$J$:$J'$:$J_1$:$J_2$\,=\,1:$-0.2$:$-0.5$:0.2 and $J$\,=\,252\,K; ii) the
$\frac13$-magnetization plateau can be understood as a product of nearly
independent polarized trimers, and iii) the effective ${\cal J}_1$-${\cal
J}_2$-${\cal J}_2'$ model shows indications for a bond nematic phase which
precedes the onset of the plateau.

{\sl Note added:} A recent NMR study~\cite{volb:yoshida16} supports our bond
nematic phase scenario below the $\frac13$-plateau.

\begin{acknowledgments}
We thank H.~Ishikawa, M.~Yoshida, T.~Yamashita, Z.~Hiroi, H.~Rosner, and
N.~Shannon for fruitful discussions.  OJ and KH were supported by the European
Research Council under the European Unions Seventh Framework Program FP7/ERC
through grant agreement n. 306447. SF and TM were supported by JSPS KAKENHI
Grants Nos. 25800225 and 23540397,
respectively. 
\end{acknowledgments}

\nocite{PW92,dft_xiang_2013,misc:mila11,misc:totsuka98,misc:mila98,misc:tonegawa00,misc:honecker01,misc:kecke07,luttinger46,TITPACK}

\setcounter{figure}{0}
\setcounter{table}{0}
\renewcommand{\thefigure}{S\arabic{figure}}
\renewcommand{\thetable}{S\arabic{table}}

\begin{widetext}

\newpage
\begin{center}
{\large
Supplemental Material for 
\smallskip
\textbf{Magnetic Behavior of Volborthite Cu$_3$V$_2$O$_7$(OH)$_2\cdot$2H$_2$O\\ 
 Determined by Coupled Trimers Rather than Frustrated Chains}

\medskip

\normalsize O. Janson, S. Furukawa, T. Momoi, P. Sindzingre, J. Richter, and K. Held}
\end{center}
\medskip

\vskip -1cm 

\begin{table}[h] \caption{\label{tab:str-dft} Comparison of total energies
yielded by nonmagnetic DFT calculations using the full-potential code
\textsc{fplo} version 9.07-41~\cite{S:FPLO}.  The respective references,
experimental method (XRD for x-ray diffraction, ND for neutron diffraction),
number of formula units in the unit cell ($Z$), the space group and the
$k$-mesh are given in columns 1-5.  $E^{\text{GGA}}$ and $E^{\text{LDA}}$
correspond to the total energies obtained using parameterizations for the
exchange and correlation potential from Refs.~\cite{S:PBE96} and \cite{S:PW92},
respectively.  The hydrogen positions are optimized with respect to total
energy, the resulting forces are below 0.01\,eV/\r{A}.
}
\begin{ruledtabular}
\begin{tabular}{r r r r r r r}
reference & method & $Z$ & sp.\ gr. & $k$-mesh & $E^{\text{GGA}}/Z$ (Hartree) & $E^{\text{LDA}}/Z$ (Hartree) \\ \hline
Ref.~\cite{S:volb:basso88}        & XRD         & 1 & $C2/m$   & 8$\times$8$\times$8 & $-7695.23828$ & $-7678.79575$ \\
Ref.~\cite{S:volb:lafontaine90}   & ND          & 1 & $C2/m$   & 8$\times$8$\times$8 & $-7695.24387$ & $-7678.80053$ \\
Ref.~\cite{S:volb:lafontaine90}   & XRD         & 1 & $C2/m$   & 8$\times$8$\times$8 & $-7695.22987$ & $-7678.79266$ \\
Ref.~\cite{S:volb:kashaev08}      & XRD         & 2 & $CC$     & 4$\times$4$\times$4 & $-7695.24642$ & $-7678.80498$ \\
Ref.~\cite{S:volb:yoshida12ncomm} & XRD, 150\,K & 2 & $C2/c$   & 4$\times$4$\times$4 & $-7695.24236$ & $-7678.80240$ \\
Ref.~\cite{S:volb:yoshida12ncomm} & XRD, 322\,K & 1 & $C2/m$   & 8$\times$8$\times$8 & $-7695.23947$ & $-7678.79349$ \\
Ref.~\cite{S:volb:ishikawa12}     & XRD         & 2 & $C2/c$   & 4$\times$4$\times$4 & $-7695.24878$ & $-7678.80835$ \\
Ref.~\cite{S:volb:ishikawa15}     & XRD         & 4 & $P2_1/c$ & 4$\times$4$\times$4 & $-7695.25191$ & $-7678.81252$ \\
\end{tabular}
\end{ruledtabular}
\end{table}

\begin{table}[htb]
\caption{\label{tab:Hstr} Crystal structure used for DFT and DFT+$U$
calculations.  We use the 50\,K data from Ref.~\cite{S:volb:ishikawa15} in the
conventional setup, i.e.\ the space group is $P2_1/c$ (14), $a$\,=\,14.41\,\r{A},
$b$\,=\,5.8415\,\r{A}, $c$\,=\,10.6489\,\r{A}, and $\beta$\,=\,95.586$^{\circ}$.
The hydrogen coordinates shown bold are optimized within GGA+$U$.  All other
atomic coordinates are taken from Ref.~\cite{S:volb:ishikawa15}.
}
\begin{ruledtabular}
\begin{tabular}{l c r r r}
atom & Wyckoff position   & $x/a$ & $y/b$ & $z/c$ \\ \hline
Cu1 & 2$d$ &$\frac12$&   0        & $\frac12$\\
Cu2 & 2$c$ & 0       &   0        & $\frac12$\\
Cu3 & 4$e$ & 0.00233 &   0.75600  & 0.24626\\
Cu4 & 4$e$ & 0.49930 &   0.74625  & 0.25280\\
V1  & 4$e$ & 0.12660 &   0.52334  & 0.49667\\
V2  & 4$e$ & 0.37297 & $-0.02131$ & 0.00304\\
O1  & 4$e$ & 0.24991 & $-0.04890$ & 0.00158\\
O2  & 4$e$ & 0.05938 &   0.00418  & 0.34368\\
\textbf{H1}  & 4$e$ & \textbf{0.12961} &   \textbf{0.01212}  & \textbf{0.34767}\\
O3  & 4$e$ & 0.55895 & $-0.00105$ & 0.34372\\
\textbf{H2}  & 4$e$ & \textbf{0.62959} &   \textbf{0.00012}  & \textbf{0.34817}\\
O4  & 4$e$ & 0.07440 &   0.50571  & 0.34159\\
O5  & 4$e$ & 0.42524 & $-0.00532$ & 0.15831\\
O6  & 4$e$ & 0.26060 &   0.50190  & 0.17836\\
O7  & 4$e$ & 0.24083 &   0.03230  & 0.31923\\
O8  & 4$e$ & 0.09912 &   0.20580  & 0.07522\\
O9  & 4$e$ & 0.60273 &   0.78700  & 0.07584\\
O10 & 4$e$ & 0.08846 &   0.73814  & 0.07118\\
O11 & 4$e$ & 0.41442 &   0.75383  & 0.42937\\
\textbf{H3}  & 4$e$ & \textbf{0.20595} &   \textbf{0.56499}  & \textbf{0.12695}\\
\textbf{H4}  & 4$e$ & \textbf{0.23942} &   \textbf{0.35162}  & \textbf{0.20575}\\
\textbf{H5}  & 4$e$ & \textbf{0.26676} &   \textbf{0.88701}  & \textbf{0.29281}\\
\textbf{H6}  & 4$e$ & \textbf{0.29312} &   \textbf{0.10203}  & \textbf{0.37325}\\
\end{tabular}
\end{ruledtabular}
\end{table}

\begin{table}[tb]
\caption{\label{tab:StJ}
Comparison of transfer integrals $t_{ij}$ (meV) evaluated using Wannier
functions for one-orbital ($x^2-y^2$, only) and two-orbital ($x^2-y^2$
and $3z^2-r^2$) models. For the latter, only the transfer integrals
between the half-filled $x^2-y^2$ are provided.  The respective band
dispersions are shown in Fig.~3 and Fig.~\ref{fig:2bandfit}. Direct
Cu..Cu distances $d_{\text{Cu..Cu}}$ are given in \r{A}).}
\begin{ruledtabular}
\begin{tabular}{l r r r}
transfer integral &
\multicolumn{1}{c}{$d_{\text{Cu..Cu}}$} &
\multicolumn{1}{c}{$t$ for one-orbital WFs} & 
\multicolumn{1}{c}{$t$ for two-orbital WFs ($x^2-y^2\leftrightarrow{}x^2-y^2$)} \\ \hline
{$t$}   &  3.053\,/\,3.058  & $-191$\,/\,$-194$ & $-198$\,/\,$-195$\\
{$t'$}  &  3.016\,/\,3.020  &  $-80$\,/\,$-84$  &  $-80$\,/\,$-85$\\
{$t_1$} &  2.922\,/\,2.923  &  $-98$\,/\,$-100$ &  $-98$\,/\,$-100$\\
{$t_2$} &  5.842\,/\,5.842  &     64\,/\,64     &     66\,/\,66\\
\end{tabular}
\end{ruledtabular}

\end{table}

\begin{figure}[!b]
\includegraphics[width=8.6cm]{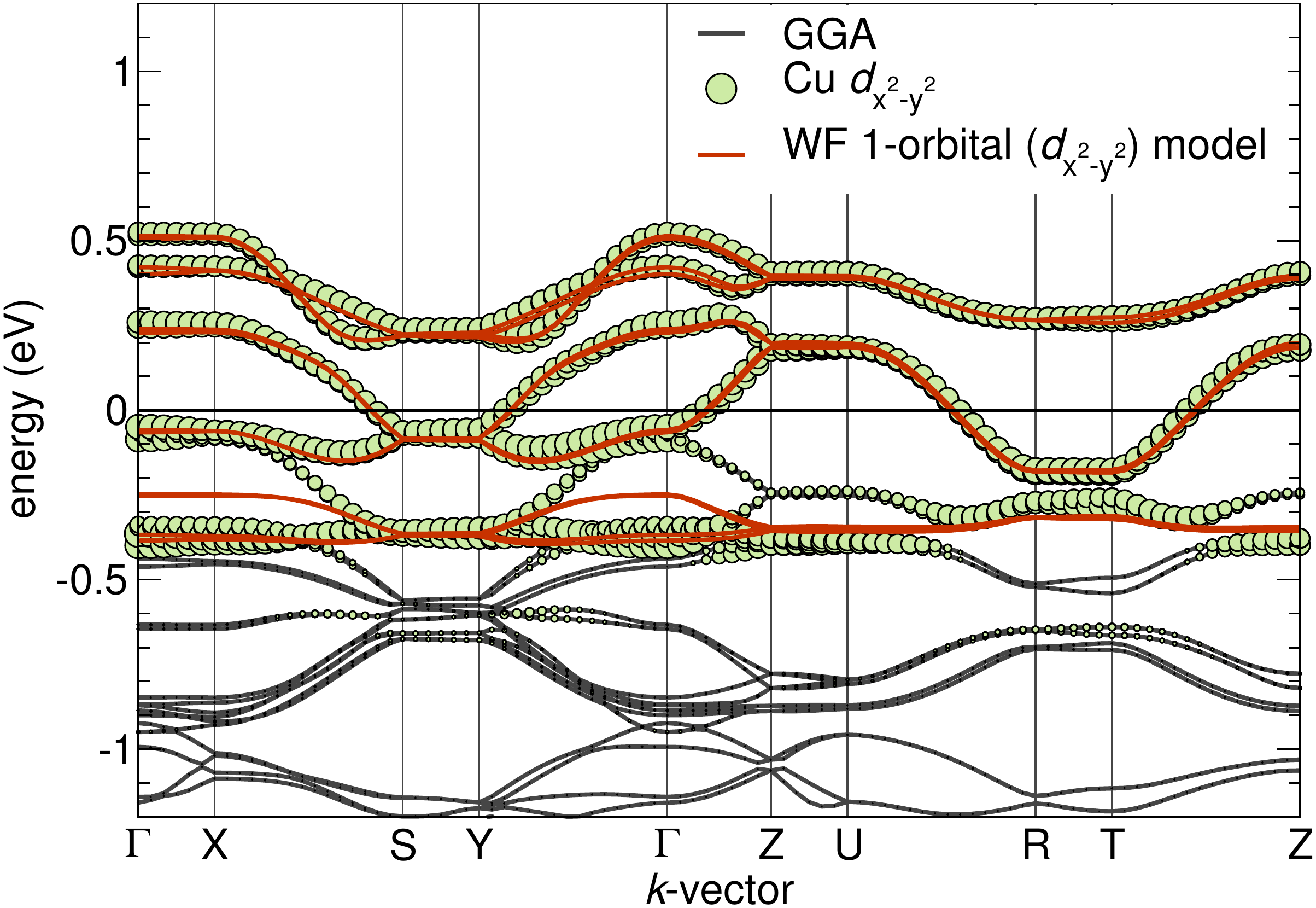}
\includegraphics[width=8.6cm]{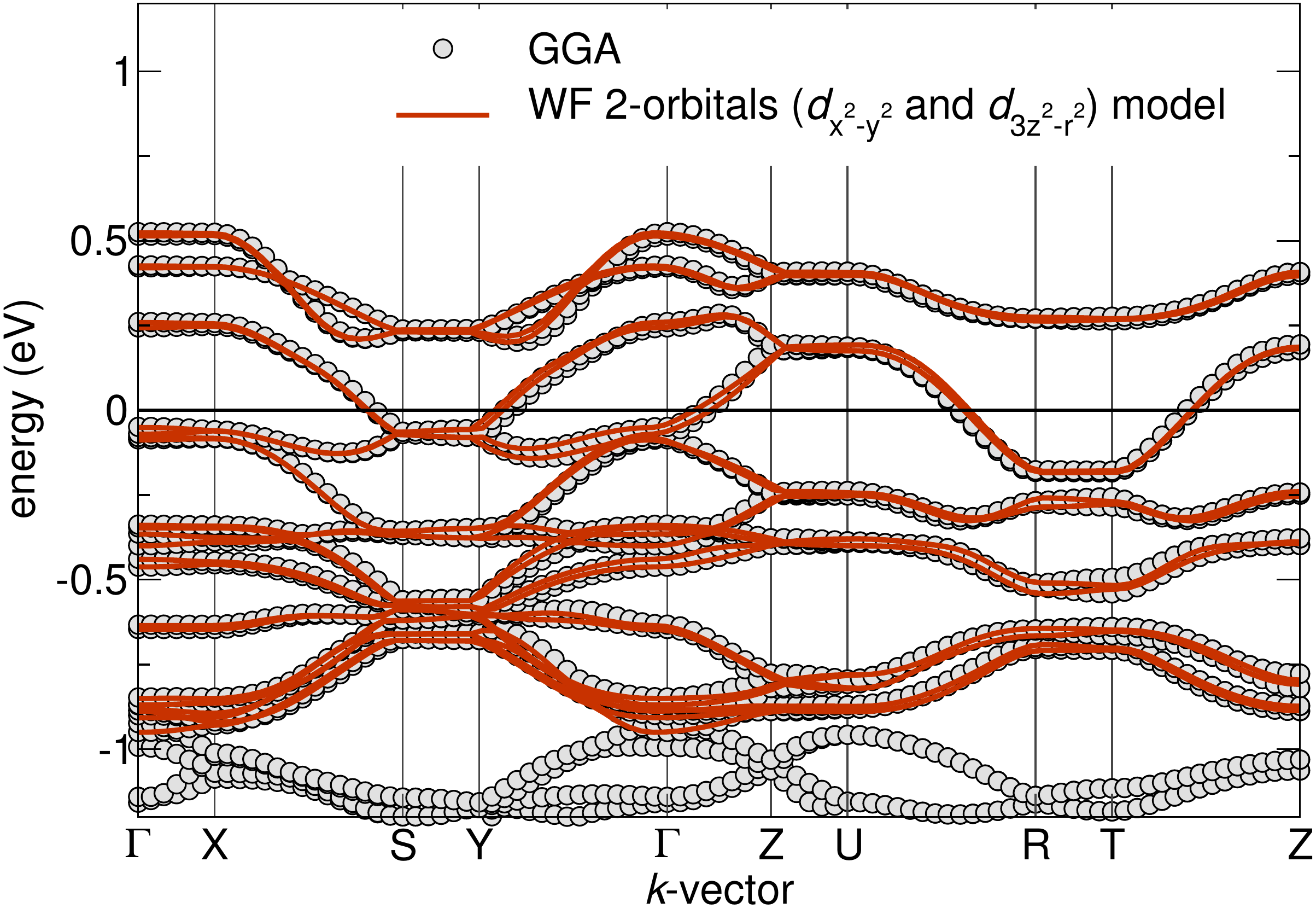}
\caption{\label{fig:2bandfit}(Color online) 
Left: GGA band structure of volborthite (solid lines) with band characters for
the Cu $d_{x^2-y^2}$ orbitals.  The dashed line indicates the projection onto
Cu-centered Wannier functions (WFs) within a one-orbital $x^2-y^2$ model.  Note
the deviations around $\sim$$-$0.4\,eV.  Right: The deviations around
$\sim$$-$0.4\,eV are remedied in a two-orbital ($x^2-y^2$ and $3z^2-r^2$)
model.  The $k$-points are: $\Gamma$\,=\,$\left(000\right)$,
X\,=\,$\left(\frac{\pi}{a}00\right)$,
M\,=\,$\left(\frac{\pi}{a}\frac{\pi}{b}0\right)$,
Y\,=\,$\left(0\frac{\pi}{b}0\right)$, Z\,=\,$\left(00\frac{\pi}{c}\right)$,
XZ\,=\,$\left(\frac{\pi}{a}0\frac{\pi}{c}\right)$,
MZ\,=\,$\left(\frac{\pi}{a}\frac{\pi}{b}\frac{\pi}{c}\right)$, and
YZ\,=\,$\left(0\frac{\pi}{b}\frac{\pi}{c}\right)$.  }
\end{figure}
\newpage
\end{widetext}

%%%%%%%%%%%%%%%%%%%%%%%%%%%%%%%%%%%%%%%%%%%%%%%%%
\section{Effective Hamiltonians}
%%%%%%%%%%%%%%%%%%%%%%%%%%%%%%%%%%%%%%%%%%%%%%%%%

In the main text, we analyze the physics at low temperatures and low fields
($T,h\equiv g\mu_B H/k_B \ll J$) by using an effective Hamiltonian of
pseudospin-$\frac12$ moments living on trimers.  Here we describe details of
the derivation of this effective Hamiltonian and our analyses of the magnon
spectra and the magnetization process of this model.  We also derive yet
another low-temperature effective Hamiltonian suitable for high fields
($T,|h-3J/2|\ll J$), which can be used to determine the high-field end
$H_\mathrm{c2}$ of the plateau phase and the saturation field $H_\mathrm{c3}$.

Our use of these effective Hamiltonians is motivated by the distinct hierarchy
of the exchange couplings $J>|J_1|>J_2,|J'|$ revealed in DFT+$U$.  We start
from the limit of large $J$, where the system decouples into trimers.  In this
limit, each isolated trimer exhibits a wide $\frac13$-magnetization plateau
over $0< h < 3J/2$.  At each endpoint of the plateau, where a level crossing
occurs in the ground state of each trimer, the total system possesses a
macroscopic degeneracy and is highly susceptible to inter-trimer couplings.
Our effective Hamiltonian is derived around each endpoint by performing
degenerate perturbation theory in terms of $J'$, $J_1$, and $J_2$.  This kind
of approach is known as strong coupling expansion \cite{S:Mila11S}, and is often
used to analyze quantum spin systems with coupled cluster structures such as
coupled dimers \cite{S:Mila11S,S:Totsuka98S,S:Mila98S} and coupled trimers
\cite{S:Tonegawa00S,S:Honecker01S}.  Our derivation of the effective Hamiltonians
goes essentially in parallel with Refs.~\cite{S:Tonegawa00S,S:Honecker01S}.

%************************************************
\subsection{Isolated trimer}
%************************************************

%############################
\begin{figure}
\begin{center}
\includegraphics[width=0.48\textwidth]{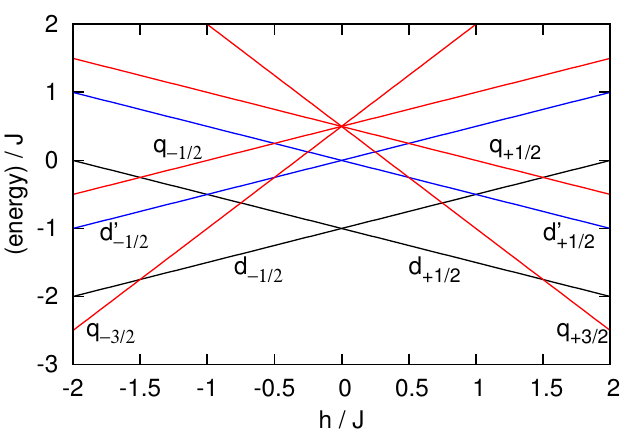}
\end{center}
\caption{\label{fig:trimer_spec}
The spectrum of an isolated trimer Hamiltonian \eqref{eq:H_trimer} as a function of the magnetic field $h$.
The eigenstates and eigenenergies are shown in Eqs.~\eqref{eq:eig_trimer} and \eqref{eq:eigener_trimer}, respectively.}
\end{figure}
%############################

We first consider an isolated trimer Hamiltonian
\begin{equation}\label{eq:H_trimer}
 H_{012} = J(\Sv_0 \cdot\ \Sv_1+\Sv_1\cdot \Sv_2) -h \sum_{j=0}^2 S^z_j,
\end{equation}
where $\Sv_1$ and $\Sv_{0,2}$ are the spin-$\frac12$ operators at the central and other sites of the trimer, respectively.
At $h=0$, because of the SU(2) symmetry and the parity symmetry around the site $1$,
the eigenstates are classified into the quadruplet $\{\ket{q_\mu}\}$, the even-parity doublet $\{\ket{d_\mu}\}$, and the odd-parity doublet $\{\ket{d'_\mu}\}$,
where $\mu$ is the eigenvalue of $\sum_{j=0}^2 S_j^z$.
Their wave functions for $\mu>0$ are given by
\begin{equation}\label{eq:eig_trimer}
\begin{split}
\ket{q_{+\frac32}} &= \ket{\ua\ua\ua},\\
\ket{q_{+\frac12}} &= \frac{1}{\sqrt{3}} \left( \ket{\ua\ua\da}+\ket{\da\ua\ua}+\ket{\ua\da\ua}\right),\\
\ket{d_{+\frac12}} &= \frac{1}{\sqrt{6}} \left( \ket{\ua\ua\da}+\ket{\da\ua\ua}-2\ket{\ua\da\ua}\right),\\
\ket{d'_{+\frac12}} &= \frac{1}{\sqrt{2}} \left( \ket{\ua\ua\da}-\ket{\da\ua\ua}\right).
\end{split}
\end{equation} Other eigenstates with $\mu<0$ are obtained by applying the spin
reversal to the above.  The Zeeman term in Eq.~\eqref{eq:H_trimer} commutes
with the Heisenberg terms, and only shifts the eigenenergies by $-h\mu$.  The
eigenenergies in the presence of a field $h$ are calculated as
\begin{equation}\label{eq:eigener_trimer}
\begin{split}
 \ket{q_\mu}:&~\frac12 J-h\mu;\\
 \ket{d_\mu}:&~-J-h\mu;\\
 \ket{d_\mu'}:&~-h\mu,
\end{split}
\end{equation}
which are plotted in Fig.~\ref{fig:trimer_spec}.
At $h=0$, the ground states are the even-parity doublet $\ket{d_{\pm\frac12}}$;
they are split for $h\ne 0$.
At $h=3J/2$, a level crossing occurs, and the ground state is replaced by the fully polarized state $\ket{q_{+\frac32}}$.
This level structure gives a wide magnetization plateau with $m/m_\mathrm{sat}=1/3$ over $0<h<3J/2$.
At the level crossing points $h=0$ and $3J/2$, the total system consisting of $N_\mathrm{t}$ trimers
possesses a macroscopic degeneracy $2^{N_\mathrm{t}}$ of ground states.
We restrict ourselves to this degenerate manifold,
and analyze the splitting of the degeneracy due to inter-trimer couplings
by deriving effective Hamiltonians.

The small Hilbert space of the isolated trimer model facilitates a direct
evaluation of its partition function, and hence its magnetic susceptibility,
which is given by:

\begin{equation}
\label{eq:chi-trimer}
\chi^{*}(\beta) = \beta \left( \frac{1}{12} +
\frac23\left({2+\exp{\frac{\beta}{2}}\ + \exp{\frac{3\beta}{2}}}\right)^{-1} \right),
\end{equation}
where $\beta$ is the inverse temperature.

%************************************************
\subsection{Effective Hamiltonian for $|m/m_{\rm sat}| \le 1/3$} %$|h| \ll J$}
%************************************************

%############################
\begin{figure}
\begin{center}
\includegraphics[width=0.48\textwidth]{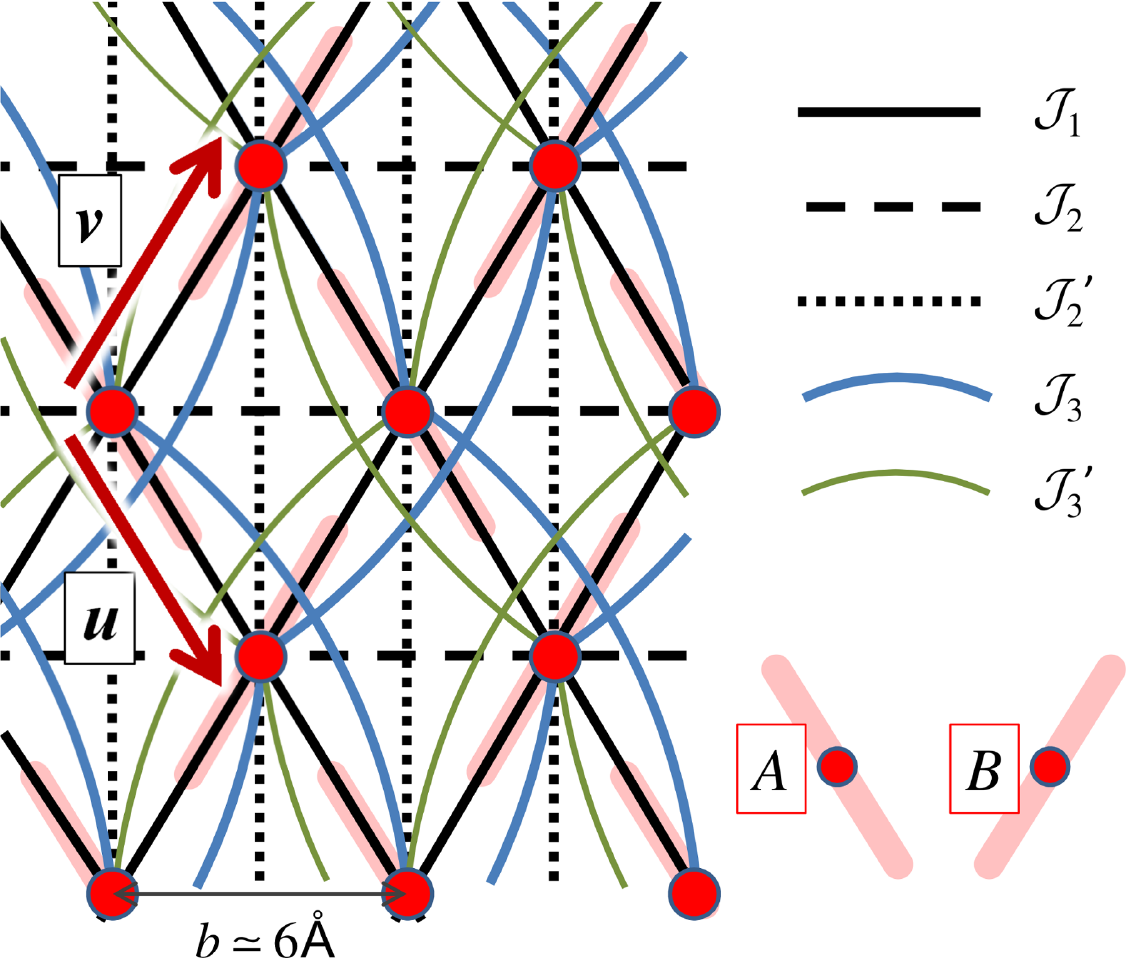}
\end{center}
\caption{\label{fig:effective}
Effective Heisenberg Hamiltonain for $|h|\ll J$.
The five largest couplings obtained in the second-order perturbation theory are displayed.
For each trimer indicated by a light red bar (with its center at $\rv$) ,
we introduce a pseudospin-$\frac12$ operator $\Tv_\rv$.
These pseudospins form a triangular lattice.
This lattice consists of two sublattices $A$ and $B$ corresponding to two different directions of trimers.
The vectors $\uv$ and $\vv$ (of lengths $|\uv|,|\vv|\simeq b\simeq 6\AA$) connect between neighboring sites on the triangular lattice.
}
\end{figure}
%############################

% [Memo on lattice constants]
% a=10.6489(1); b=5.8415(1)
% |u|=|v|=sqrt(a^2+b^2)/4=6.073

%############################
\begin{table*}
\caption{\label{table:estimates}
Nonzero coupling constants $\Jcal_{\Delta\rv,X}$ in the effective Hamiltonian \eqref{eq:Heff2}.
The vectors $\uv$ and $\vv$ are defined in Fig.~\ref{fig:effective}.
Only for $\Jcal_3$ and $\Jcal_3'$, the relative vector $\Delta\rv$ depends on the sublattice $X$,
as indicated in the corresponding rows.
The first- and second-order perturbative estimates (third and fourth columns)
are calculated by substituting Eq.~\eqref{eq:J_volb} into Eqs~\eqref{eq:Jeff1} and \eqref{eq:Jeff2}.
We also consider Models I and II shown in the fifth and sixth columns,
where the leading three and four couplings in the second-order model are taken into account, respectively;
these simplified models help to understand the essential physics arising from the leading couplings.
For each of these models, one- and two-magnon condensation points,
$h_\mathrm{c1}^{(n)}=g\mu_B H_\mathrm{c1}^{(n)}/k_B$ with $n=1,2$, are presented.
Here, $g=2.151$ obtained in the main text is used to convert $h_\mathrm{c1}^{(n)}$ into $H_\mathrm{c1}^{(n)}$.
In the four models, the one-magnon spectrum has the minimum at incommensurate wave vectors $\kv b=\pm(Q,0)$,
and the value of $Q/(2\pi)$ is also presented.
}
\begin{ruledtabular}
\begin{tabular}{ccrrrr}
 & relative vectors $\Delta\rv$ & 1st-order & 2nd-order  & Model I & Model II \\
  \hline
${\cal J}_1$ & $\uv,\vv$ & $-44.8$ K & $-34.9$ K & $-34.9$ K & $-34.9$ K\\
${\cal J}_2$ & $\uv+\vv$ & 44.8 K & 36.5 K & 36.5 K & 36.5 K \\
${\cal J}_2^{\prime}$ & $-\uv+\vv$ &  & 6.8 K & 6.8 K & 6.8 K \\
${\cal J}_3$ & $(2\uv,A),~(2\vv,B)$ & & 4.6 K &  & 4.6 K \\
${\cal J}_3^{\prime}$ & $(2\vv,A),~(2\uv,B)$ & & 1.7 K &  &  \\
${\cal J}_4$ & $2\uv+\vv,\uv+2\uv$ &  & 1.7 K & & \\
${\cal J}_5$ & $2(\uv+\vv)$ &  & $-1.3$ K& & \\
\hline
$h_\mathrm{c1}^{(1)}~(H_\mathrm{c1}^{(1)})$ && 22.4 K (15.5 T) & 35.0 K (24.2 T) & 19.9 K (13.8 T) & 27.2 K (18.9 T)\\
$Q/(2\pi)$ && 0.333 & 0.370 & 0.341 & 0.360\\
$h_\mathrm{c1}^{(2)}~(H_\mathrm{c1}^{(2)})$ && 22.4 K (15.5 T) & 35.0 K (24.2 T) & 25.7 K (17.8 T) & 29.0 K (20.1 T)
\end{tabular}
\end{ruledtabular}
\end{table*}
%############################

Here we derive the effective Hamiltonian for the range $|m/m_{\rm sat}|\le 1/3$.
We label each trimer by the position $\rv$ of its central site.
These positions form a triangular lattice as shown in Fig.~\ref{fig:effective}.
This lattice consists of two sublattices $A$ and $B$ corresponding to two different directions of trimers.

For $|h|\ll J$, we use the lowest-energy doublet $\ket{d_{\pm\frac12}}_\rv$ at $h=0$
as the local basis on each trimer $\rv$.
Using these states, we introduce a pseudospin-$\frac12$ operator
\begin{equation}
 \Tv_\rv
 = \left( \ket{d_{+\frac12}}_\rv, \ket{d_{-\frac12}}_\rv \right) \frac{\bm{\sigma}}{2}
 \begin{pmatrix} _\rv\bra{d_{+\frac12}} \\ _\rv\bra{d_{-\frac12}} \end{pmatrix},
\end{equation}
where $\bm{\sigma}=(\sigma^x,\sigma^y,\sigma^z)$ are the Pauli matrices.

The first-order effective Hamiltonian is derived by projecting the inter-trimer interactions
onto the degenerate manifold $V_0 = \bigotimes_\rv \mathrm{Span}(\{\ket{d_{\pm\frac12}}_\rv\})$.
The obtained Hamiltonian is a Heisenberg model of pseudospins $\{\Tv_\rv\}$ on the triangular lattice
with spatially anisotropic nearest-neighbor couplings
\begin{equation}\label{eq:Jeff1}
{\cal J}_1^{\rm (1st)} = \frac{2}{9}(2 J_1-J^\prime),~~
{\cal J}_2^{\rm (1st)} = \frac{8}{9} J_2,
\end{equation}
as shown in Fig.~\ref{fig:effective}.

To calculate the second-order effective Hamiltonian, we take into account various processes
where a state in $V_0$ is virtually promoted to an excited state outside $V_0$, and then comes back to $V_0$ by the operations of the inter-trimer interactions.
Such virtual processes give rise to various further-neighbor interactions between pseudospins.
The resulting Hamiltonian is a Heisenberg model
\begin{equation}\label{eq:Heff2}
 H_\mathrm{eff}=\sum_\rv \sum_{\Delta\rv} {\cal J}_{\Delta\rv,X_\rv}\Tv_\rv \cdot \Tv_{\rv+\Delta\rv} -h \sum_{\rv} T^z_\rv,
\end{equation}
where $X_\rv=A,B$ indicates the sublattice which $\rv$ belongs to.
The coupling constants ${\cal J}_{\Delta\rv,X}$ show seven nonzero different values as listed in Table~\ref{table:estimates},
and their second-order expressions are given by
\begin{equation}\label{eq:Jeff2}
\begin{split}
{\cal J}_1^{\rm (2nd)} =& \frac{2}{9}(2 J_1-J^\prime)
+\frac{211 {J_1}^2 + 48 J_1 J^\prime - 118 {J^\prime}^2}{1620 J} \\
&+ \frac{8 J_2 (-4 J_1 + 5 J^\prime)}{243 J}, \\
{\cal J}_2^{\rm (2nd)} =& \frac{8}{9} J_2 - \frac{{J_2}^2}{81 J}
-\frac{2 (-2 J_1 + J^\prime)(-13 J_1 + 8 J^\prime)}{243 J}, \\
{\cal J}_2^{\prime {\rm (2nd)}} =& \frac{2(-2 J_1 + J^\prime) (-5 J_1 - 8 J^\prime)}{243 J},\\
{\cal J}_3^{\rm (2nd)} =& \frac{2 (-5 J_1 + 4 J^\prime) (-J_1 - 4 J^\prime)}{243 J},\\
{\cal J}_3^{\prime {\rm (2nd)}} =& \frac{5 (-2 J_1 + J^\prime)^2}{486 J},\\
{\cal J}_4^{\rm (2nd)} =& \frac{8 J_2 (-4 J_1 + 5 J^\prime)}{243 J},\\
{\cal J}_5^{\rm (2nd)} =& -\frac{32 {J_2}^2}{243 J}.
\end{split}
\end{equation}
Most of these interactions do not depend on the sublattice $X$,
and have the translational invariance of the triangular lattice.
Only $\Jcal_3$ and $\Jcal'_3$ depend on the sublattice $X$, and double the unit cell of the effective model when $\Jcal_3\ne\Jcal_3'$;
the primitive vectors of the system then change to $\pm\uv+\vv$. Using
\begin{equation}\label{eq:J_volb}
J:J':J_1:J_2=1:-0.2:-0.5:0.2,~~J\simeq 252~\mathrm{K},
\end{equation}
obtained in the main text,
the effective coupling constants are calculated as in the columns ``1st-order'' and ``2nd-order'' in Table~\ref{table:estimates}.
The five largest couplings in the second-order model are displayed in Fig.~\ref{fig:effective}.

%************************************************
\subsection{Magnon spectra}
%************************************************

Using the effective model $H_\mathrm{eff}$ in Eq.~\eqref{eq:Heff2},
we here calculate the spectra of one- and two-magnon excitations which lower the total magnetization by one and two, respectively,
from the $\frac13$-magnetization plateau state.
The plateau state corresponds to the fully polarized state of pseudospins
$\bigotimes_\rv \ket{d_{+\frac12}}_\rv=:\ket{\mathrm{vac}}$, which we view as the magnon vacuum in the following.
Then, $T^\pm_\rv:=T^x_\rv\pm i T^y_\rv$ play the roles of  magnon annihilation and creation operators,
and the magnon occupation number at the site $\rv$ is given by $n_\rv:=\frac12-T_\rv^z$.
Using these operators, the Hamiltonian is rewritten as
\begin{equation}\label{eq:Heff_Hmag}
 H_\mathrm{eff}=N_\mathrm{t} \left( \frac14 \Jcal -\frac{h}{2} \right) + h \sum_\rv n_\rv + H_\mathrm{mag},
\end{equation}
where
\begin{align}
 \Jcal &= \frac1{N_\mathrm{t}} \sum_\rv \sum_{\Delta\rv} \Jcal_{\Delta\rv,X_\rv} \notag\\
 &=2\Jcal_1+\Jcal_2+\Jcal_2'+\Jcal_3+\Jcal_3'+2 \Jcal_4+\Jcal_5,\\
 H_\mathrm{mag} &= -\Jcal \sum_\rv n_\rv \notag\\
 &\hspace{-0.6cm}+ \sum_\rv \sum_{\Delta\rv} {\cal J}_{\Delta\rv,X_\rv} \left[
 \frac12 \left( T^+_\rv T^-_{\rv+\Delta\rv} + \mathrm{h.c.} \right) + n_\rv n_{\rv+\Delta\rv} \right].\label{eq:H_mag}
\end{align}
The first term in Eq.~\eqref{eq:Heff_Hmag} is the energy of the vacuum $\ket{\mathrm{vac}}$.
The second term is the Zeeman term, which plays the role of a magnon chemical potential.
The third term $H_\mathrm{mag}$ contains magnon kinetic and interaction terms, and does not depend on $h$.
Below we calculate the $n$-magnon spectrum of $H_\mathrm{mag}$,  and in particular determine the lowest energy $E^{(n)}$ in it.
We then find from Eq.~\eqref{eq:Heff_Hmag}
that the minimum energy cost for creating $n$ magnons from the vacuum is given by $E^{(n)}+h n$.
This energy reaches zero at $h=h_\mathrm{c1}^{(n)}=-E^{(n)}/n$, which corresponds to an $n$-magnon condensation point.

While the second-order effective model is expected to describe well the properties of the original model with Eq.~\eqref{eq:J_volb} for $T,h\ll J$,
it contains small couplings of magnitudes less than $2$ K, which may be influenced easily
by potential parameter changes or inclusion of small further-neighbor couplings in the original model,
which are not estimated in the DFT+U calculation.
We therefore consider also Models I and II shown in Table~\ref{table:estimates},
where the leading three and four couplings in the second-order model are taken into account, respectively.
These simplified models help to understand the essential physics arising from the leading effective couplings.

%############################
\begin{figure*}
\begin{center}
\includegraphics[width=0.4\textwidth]{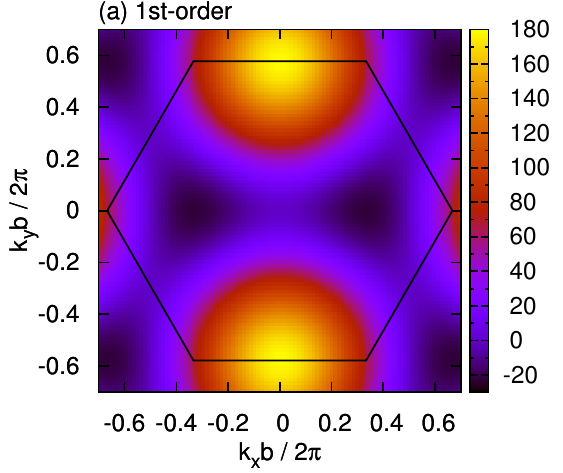}
\includegraphics[width=0.4\textwidth]{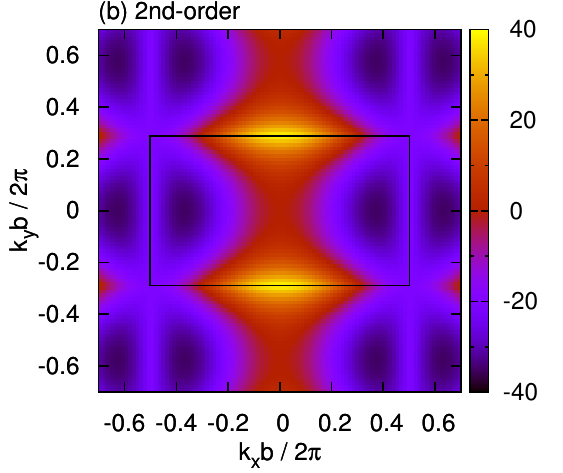}
\includegraphics[width=0.4\textwidth]{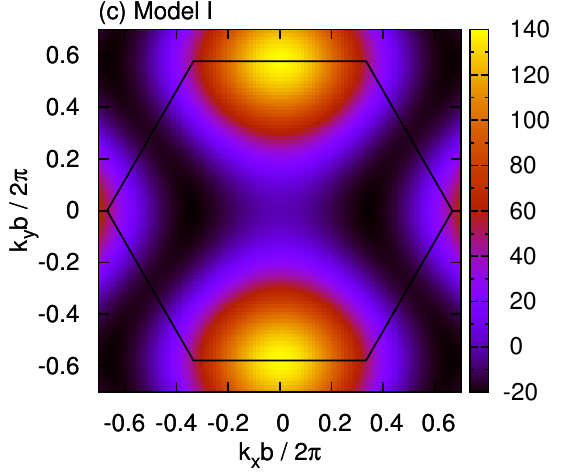}
\includegraphics[width=0.4\textwidth]{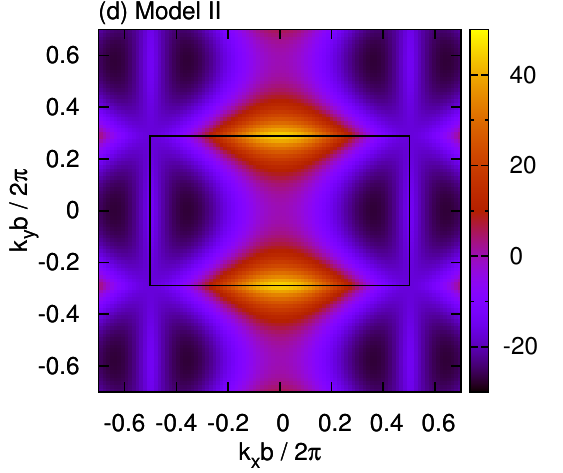}
\end{center}
\caption{\label{fig:magspec1}
One-magnon spectra of $H_\mathrm{mag}$ in Eq.~\eqref{eq:H_mag} for the four models shown in Table~\ref{table:estimates}.
We plot $\epsilon(\kv)$ in Eq.~\eqref{eq:epsilon_k} for (a) and (c) ,
and $\epsilon_-(\kv)$ in Eq.~\eqref{eq:epsilon_pm_k} for (b) and (d).
The color indicates an energy in units of Kelvin.
A hexagon or a square indicates the first Brillouin zone.
The wave numbers, $k_x$ and $k_y$, are defined along the horizontal and vertical directions of Fig.~\ref{fig:effective}, respectively.
}
\end{figure*}
%############################

%############################
\begin{figure*}
\begin{center}
\includegraphics[width=0.4\textwidth]{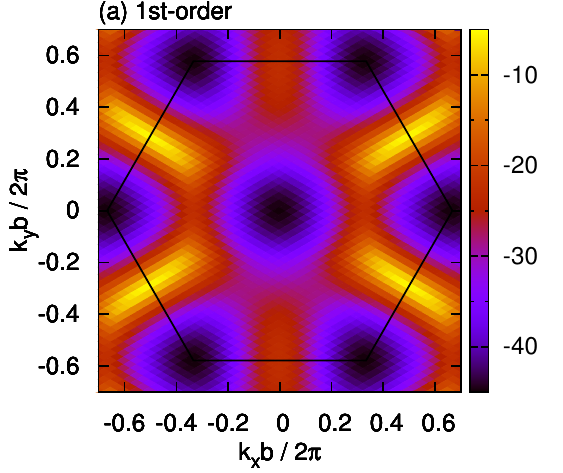}
\includegraphics[width=0.4\textwidth]{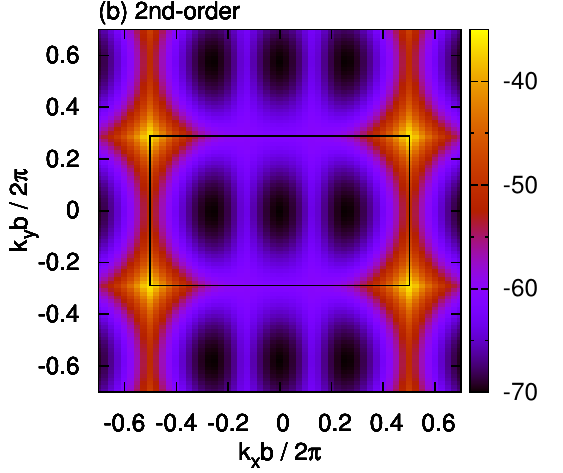}
\includegraphics[width=0.4\textwidth]{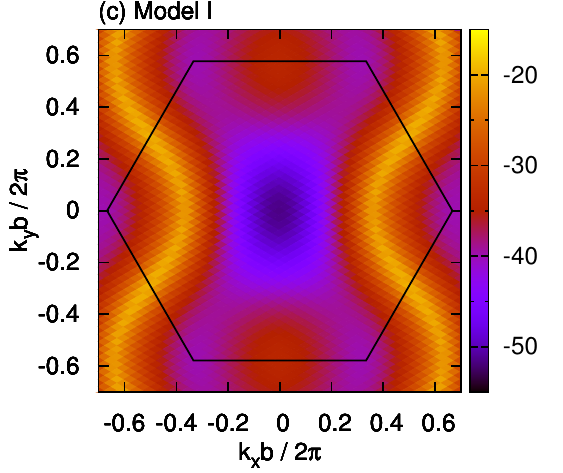}
\includegraphics[width=0.4\textwidth]{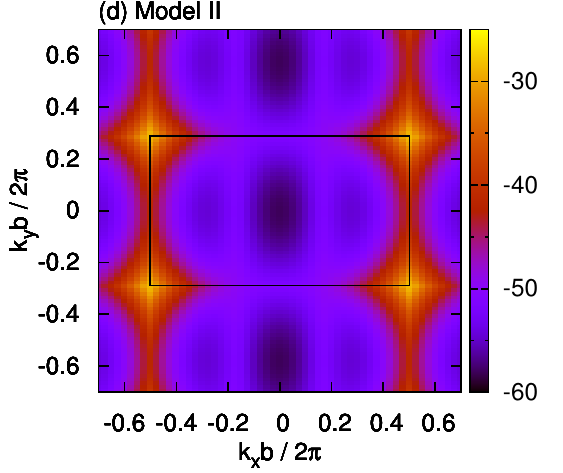}
\end{center}
\caption{\label{fig:magspec2}
Two-magnon spectra of $H_\mathrm{mag}$ in Eq.~\eqref{eq:H_mag} for the four models shown in Table~\ref{table:estimates}.
}
\end{figure*}
%############################

We first analyze one-magnon spectra of $H_\mathrm{mag}$.
When $\Jcal_3=\Jcal_3'$, the effective model has the translational invariance of the triangular lattice.
In this case, one-magnon state forms a single band
\begin{equation}\label{eq:epsilon_k}
 \epsilon (\kv) = -\Jcal + \sum_{\Delta\rv} \Jcal_{\Delta\rv} \cos (\kv\cdot\Delta\rv).
\end{equation}
When $\Jcal_3\ne \Jcal_3'$, the unit cell is doubled,
and the one-magnon bands are obtained by diagonalizing the $2\times 2$ matrix
\begin{equation}
 M(\kv) = \epsilon_0 (\kv) I + \bm{J} (\kv) \cdot \bm{\sigma},
\end{equation}
where $I$ is the identity matrix, and
\begin{equation}
\begin{split}
 \epsilon_0 (\kv) &= -\Jcal
 +\Jcal_2\cos (k_u+k_v) + \Jcal_2' \cos(k_u-k_v) \\
 &~~~+ \frac12 (\Jcal_3+\Jcal_3') \left[ \cos(2k_u)+\cos(2k_v) \right] \\
 &~~~+ \Jcal_5\cos(2k_u+2k_v), \\
 J^x(\kv)&= \Jcal_1 \left( \cos k_u + \cos k_v \right) \\
 &~~~ + \Jcal_4 \left[ \cos(2k_u+k_v) + \cos(k_u+2k_v) \right], \\
 J^y(\kv)&= 0,\\
 J^z(\kv)&= \frac12 (\Jcal_3-\Jcal_3') \left[ \cos(2k_u)-\cos(2k_v) \right]
\end{split}
\end{equation}
with $k_u=\kv\cdot\uv$ and $k_v=\kv\cdot\vv$.
The two bands are calculated as
\begin{equation}\label{eq:epsilon_pm_k}
 \epsilon_\pm (\kv) = \epsilon_0 (\kv) \pm |\bm{J}(\kv)|.
\end{equation}

The (lower) one-magnon band $\epsilon(\kv)$ or $\epsilon_-(\kv)$ is plotted for the four models in Fig.~\ref{fig:magspec1}.
In all the cases, the spectrum has the minimum energy $E^{(1)}$ at incommensurate wave vectors $\kv b=\pm (Q,0)$.
The one-magnon condensation point $h_\mathrm{c1}^{(1)}=-E^{(1)}$ and the incommensurate wave number $Q/(2\pi)$ are presented in Table~\ref{table:estimates}.
Remarkably, the band $\epsilon(\kv)$ for Model I shows a line of nearly degenerate minima extending roughly along the vertical direction.
This implies suppression of magnon hopping (and hence relative enhancement of magnon interactions) along this direction.

We next analyze two-magnon spectra.
To this end, we introduce the two-magnon basis \cite{S:Kecke07S}, which is given by
\begin{equation}\label{eq:2magbasis}
 \ket{\Delta\rv,\kv} = \frac{1}{\sqrt{N_\mathrm{t}}} \sum_\rv e^{i\kv\cdot\rv} T^-_\rv T^-_{\rv+\Delta\rv} \ket{\mathrm{vac}}
\end{equation}
when $\Jcal_3=\Jcal_3'$, and by
\begin{equation}\label{eq:2magbasisX}
 \ket{X,\Delta\rv,\kv} = \frac{1}{\sqrt{N_\mathrm{t}/2}} \sum_{\rv\in X} e^{i\kv\cdot\rv} T^-_\rv T^-_{\rv+\Delta\rv} \ket{\mathrm{vac}}
\end{equation}
with $X=A,B$ when $\Jcal_3\ne\Jcal_3'$.
Here, the magnon relative vector $\Delta\rv$ is chosen in the range $|\Delta\rv|/b \lesssim O(10^2)$
in such a way that the double counting of $\pm\Delta\rv$ which lead to the same state is avoided.
The matrix elements of $H_\mathrm{mag}$ are given by
\begin{align}
 &\bra{\Delta\rv,\kv} H_\mathrm{mag} \ket{\Delta\rv',\kv} \notag\\
 &= \sum_\rv e^{i\kv\cdot\rv} \bra{\mathrm{vac}} T_{\bm{0}}^+ T_{\Delta\rv}^+ H_\mathrm{mag} T_\rv^- T_{\rv+\Delta\rv'}^- \ket{\mathrm{vac}},\\
 &\bra{X,\Delta\rv,\kv} H_\mathrm{mag} \ket{X',\Delta\rv',\kv} \notag\\
 &= \sum_{\rv\in X'} e^{i\kv\cdot(\rv-\uv \delta_{XB} )} \notag\\
 &~~~~\times\bra{\mathrm{vac}} T_{\uv\delta_{XB}}^+ T_{\uv\delta_{XB}+\Delta\rv}^+ H_\mathrm{mag} T_\rv^- T_{\rv+\Delta\rv'}^- \ket{\mathrm{vac}}
\end{align}
for the bases in Eqs.~\eqref{eq:2magbasis} and \eqref{eq:2magbasisX}, respectively.
We note that the dependence on $N_\mathrm{t}$ has dropped in these expressions,
and we are effectively treating an infinite system (an error can only arise from the finite cutoff for $|\Delta\rv|$).
By performing Lanczos diagonalization for such matrices,
we have calculated the lowest eigenenergy for each $\kv$,
which is plotted in Fig.~\ref{fig:magspec2}.
In (a) the first-order and (b) second-order models,
we find nearly degenerate minima at $\kv b=(0,0)$ and $\pm (2Q,0)$ with energy $E^{(2)} \approx 2E^{(1)}$;
these can be interpreted as two independent magnons, each with the one-magnon lowest energy $E^{(1)}$.
In (c) Model I and (d) Model II, by contrast, a single minimum at $\kv b=(0,0)$ with energy $E^{(2)}<2E^{(1)}$ is formed,
implying a significant effect of attractive interactions.
In (c) Model I, in particular, two magnons acquire an appreciable binding energy $2E^{(1)}-E^{(2)}=11.6$ K,
which is likely to be due to aforementioned suppression of magnon hopping along the vertical direction.
The two-magnon condensation point $h_\mathrm{c1}^{(2)}=-E^{(2)}/2$ is presented in Table~\ref{table:estimates}.
The relation $h_\mathrm{c1}^{(1)}<h_\mathrm{c1}^{(2)}$ seen in Models I and II
indicates that bimagnon condensation leading to a bond nematic order
occurs with lowering the field $h$ from the plateau phase.
We have also calculated $3$- and $4$-magnon spectra (not shown),
finding no indication of multimagnon condensation with $h_\mathrm{c1}^{(n)}>h_\mathrm{c1}^{(1)},h_\mathrm{c1}^{(2)}~(n=3,4)$.

%************************************************
\subsection{Exact diagonalization}
%************************************************

In the main text, we present the exact diagonalization result for the magnetization process of
the second-order effective model in Eq.~\eqref{eq:Heff2}.
Here we explain the cluster shapes used for this study.
A finite-size cluster is specified by two vectors $\bm{L}_j=L_{uj}\uv+L_{vj}\vv$ [$j=1,2;~L_{uj},(L_{uj}+L_{vj})/2\in\mathbb{Z}$],
which set the periodic boundary conditions $\Tv_\rv\equiv \Tv_{\rv+\bm{L}_j}$.
The number of pseudospins in the system is given by $N_\mathrm{t}=|L_{u1}L_{v2}-L_{v1}L_{u2}|$.
We have used the following clusters specified by $(L_{u1},L_{v1})$ and $(L_{u2},L_{v2})$:
\begin{equation}
\begin{split}
 %&N_\mathrm{t}=22:~(5,1),~(-2,4); \\
 &N_\mathrm{t}=24:~(5,3),~(-3,3); \\
 &N_\mathrm{t}=26:~(5,1),(-1,5); \\
 &N_\mathrm{t}=30:~(5,-1),(0,6).
\end{split}
\end{equation}
The reason for these choices is as follows.
In finite-size clusters with periodic boundary conditions,
two distinct interactions $\Tv_\rv\cdot \Tv_{\rv+\Delta\rv}$ and $\Tv_\rv\cdot \Tv_{\rv+\Delta\rv'}$
fall into the identical one if $\Delta\rv-\Delta\rv'=m_1\bm{L}_1+m_2\bm{L}_2~(m_1,m_2\in\mathbb{Z})$.
In the above choices, where $\bm{L}_1$ and $\bm{L}_2$ are sufficiently long, such a situation can be avoided.
Furthermore, $\bm{L}_1$ and $\bm{L}_2$ form nearly $120^\circ$ to each other in the above choices,
making the cluster shape as isotropic as possible.
For example, $120^\circ$ rotation of $(L_{u1},L_{v1})=(5,1)$ gives $(-L_{v1},L_{u1}-L_{v1})=(-1,4)$.
However, the latter vector connects between different sublattices,
and thus we shift it slightly and set $(L_{u2},L_{v2})=(-1,5)$ to obtain the $N_\mathrm{t}=26$ cluster.
Exact diagonalization of the effective model was performed using TITPACK ver. 2~\cite{S:TITPACK}.

%************************************************
\subsection{Effective Hamiltonian for $1/3 \le m/m_{\rm sat} \le 1$} %$|h-3J/2|\ll J$}
%************************************************

For $|h-3J/2|\ll J$, we can derive yet another effective model valid in the range $1/3 \le m/m_{\rm sat} \le 1$,
by using $\ket{q_{+\frac32}}_\rv$ and $\ket{d_{+\frac12}}_\rv$ as the local basis on each trimer $\rv$.
Using these states, we introduce a pseudospin-$\frac12$ operator
\begin{equation}
 \tilde{\Tv}_\rv
 = \left( \ket{q_{+\frac32}}_\rv, \ket{d_{+\frac12}}_\rv \right) \frac{\bm{\sigma}}{2}
 \begin{pmatrix} _\rv\bra{q_{+\frac32}} \\ _\rv\bra{d_{+\frac12}} \end{pmatrix}.
\end{equation}
The all down state of the pseudospins ($\tilde{T}^z_\rv=-1/2$) corresponds to the $\frac13$-magnetization plateau of the original model;
the all up state  ($\tilde{T}^z_\rv=+1/2$) corresponds to the saturation.
The first-order effective Hamiltonian is derived in a way similar to the previous case,
and has the form of an XXZ model on the triangular lattice
with spatially anisotropic couplings.
The Hamiltonian is given by
\begin{equation}\label{eq:Hteff}
\begin{split}
 \tilde{H}_\mathrm{eff}=\sum_\rv \sum_{\Delta\rv}
 &\bigg[ \Jcal_{\Delta\rv}^{xy} (\Tt_\rv^x \Tt_{\rv+\Delta\rv}^x+\Tt_\rv^y \Tt_{\rv+\Delta\rv}^y)\\
 &+\Jcal_{\Delta\rv}^z \Tt_\rv^z \Tt_{\rv+\Delta\rv}^z \bigg]- \tilde{h} \sum_{\rv} \Tt^z_\rv,
\end{split}
\end{equation}
where
\begin{equation}
\begin{split}
 &\Jcal_1^{xy}:=\Jcal_{\uv}^{xy}=\Jcal_{\vv}^{xy}=\frac16(J_1-2J'),\\
 &\Jcal_1^z:=\Jcal_{\uv}^z=\Jcal_{\vv}^z=\frac1{36} (J_1+4J'),\\
 &\Jcal_2^{xy}:=\Jcal_{\uv+\vv}^{xy}=\frac13 J_2,\\
 &\Jcal_2^z:=\Jcal_{\uv+\vv}^z=\frac1{18}J_2,\\
 &\tilde{h}=h-h_0,\ \ \ h_0=\frac32 J+\frac1{18} (5J_1+5J_2+11J').
\end{split}
\end{equation}
Using Eq.~\eqref{eq:J_volb}, the parameters in this model are calculated as
\begin{equation}\label{eq:J_Hteff}
\begin{split}
 &(\Jcal_1^{xy}, \Jcal_1^z,\Jcal_2^{xy},\Jcal_2^z)=(-4.2,-9.1, 16.8, 2.8) ~\mathrm{K},\\
 &h_0=326~\mathrm{K}.
\end{split}
\end{equation}
Because of the spin-reversal symmetry of the XXZ Hamiltonian,
the magnetization process of the original model is symmetric about $h=h_0$
at this order of perturbation theory.

We determine the saturation field $\tilde{h}_\mathrm{s}$ of the effective model \eqref{eq:Hteff} by analyzing the single-magnon instability of the saturated state.
One-magnon excitation band above the saturated state is calculated as
\begin{equation}
\begin{split}
 \tilde{\epsilon}(\kv)
 = &\tilde{h}-2\Jcal_1^z-\Jcal_2^z+\Jcal_1^{xy}(\cos k_u + \cos k_v) \\
 &+ \Jcal_2^{xy}\cos (k_u+k_v).
\end{split}
\end{equation}
When $0<-\Jcal_1^{xy}<2J_2^{xy}$, this has the minimum at $\kv b=\pm (Q,0)$ with $Q=2\arccos \frac{|\Jcal_1^{xy}|}{2\Jcal_2^{xy}}$
and the saturation field $\tilde{h}_\mathrm{s}$ is determined as the field $\tilde{h}$ at which this minimum reaches zero, hence
\begin{equation}
 \tilde{h}_\mathrm{s}= 2\Jcal_1^z+\Jcal_2^z+\Jcal_2^{xy}+\frac{(\Jcal_1^{xy})^2}{2\Jcal_2^{xy}}.
\end{equation}

Using Eq.~\eqref{eq:J_Hteff}, the high-field end $h_\mathrm{c2}$ of the plateau and the saturation field $h_\mathrm{c3}$ of the original model are calculated as
\begin{equation}
\begin{split}
 &(h_\mathrm{c2},h_\mathrm{c3})=(h_0-\tilde{h}_\mathrm{s},h_0+\tilde{h}_\mathrm{s})=(324,328)~\mathrm{K} \\
 &[(H_\mathrm{c2},H_\mathrm{c3})=(224,227)~\mathrm{T}] .
\end{split}
\end{equation}
These fields agree well with the field at which rapid increase of $m/m_\mathrm{sat}$ is found in the exact diagonalization result of the original model.
This rapid increase in the short field width $H_\mathrm{c3}-H_\mathrm{c2}=2\tilde{h}_\mathrm{s}=2.6$ T
arises from the rather small coupling constants [Eq.~\eqref{eq:J_Hteff}] in the effective XXZ model \eqref{eq:Hteff}.
For $H_\mathrm{c2}<H<H_\mathrm{c3}$, the system is expected to exhibit an incommensurate magnetic order with a wave vector $Q/(2\pi)=0.46$.


\begin{thebibliography}{54}
\bibitem [{foo()}]{footnote1}%
  \BibitemOpen
  \href@noop {} {}\bibinfo {note} {In a magnetic field, the transverse
  component of a bond nematic order can be decomposed into the $x^2$$-$$y^2$
  component given by $D_{ij}\propto\cos{2\theta}$ and the $xy$ component
  $\propto{}\sin{2\theta}$, where $\theta$ is the rotation around the
  $z$-axis~\cite{eff:shannon06}. Symmetry breaking will pick one specific
  $\theta$, which we set to $\theta$\,=\,0 ($\theta$\,=\,$\frac{\pi}{2}$
  follows on the neighboring sites). Hence the $xy$-component vanishes, and
  $D_{ij}$ acquires alternating signs on the neighboring bonds
  (Fig.~\ref{fig:nematic}, right).}\BibitemShut {Stop}%
\end{thebibliography}
\end{document}